# Digital Enablers Of Construction Project Governance

Paolo Eugenio Demagistris[1], Sandro Petruzzi[2], Rodolfo Pampaloni[5], Milan Šmigić[4], Alberto De Marco[1], Waseem Khan[3], and Filippo Maria Ottaviani[1]

[1]*Politecnico di Torino*
[2]*Università degli Studi di Torino*
[3]*Putra Business School*
[4]*Centar za upravljanje projektima CPM*
[5]*Studio legale Avvocato Pampaloni*

## Abstract

Construction project governance relies on agreements between the actors along the construction industry value chain. The mutual obligations arising from these contracts rely on timely monetary trans- actions. Despite the advantages of automation in payment systems and improved access to digital progress data, several payment applications rely nonetheless on inefficient and time-consuming procedures and documentation. This study examines the present technological advancements that can lead to fix this problem. A smart contract-based approach is ideal for managing construction progress payments that support autonomous process, it fills the gap between payments and project site progress evaluations. This article offers a way for automat- ing construction payments by formalizing smart contracts execution on a decentralized block-chain-based system.

## 1 Introduction

The implementation of contracts governing the conduct of stakeholders by keeping them liable by obligations, restrictions, liabilities and commitments is very important for the resolution of payment difficulties in building projects. It provides the foundation for information, claims and payments management, and is there- fore essential to project success. Construction contract execution, on the other hand, is a complicated procedure fraught with difficulties, such as payment de- lays, Davison and Sebastian [2009]. In circumstances like defaulting on payment due to a lack of cash or arguments about unacceptable work quality, a contract is considered to have been breached or interrupted. The uncertainty of terminology in identifying responsibility, authority, and prohibitions mentioned in the contracts exacerbates these issues. Furthermore, the contract management process is slow because decision-making requires consensus among parties. Multiple layers of agreement from stakeholders from many agencies precede the consensus; hence, an unchanging audit trail must be developed to avoid future disagreements. Measures to improve construction contract management were offered in previous studies. Standard construction contracts, on the other hand, are focused on improving the contract structure and are nevertheless difficult to comprehend by those who are not lawyers. Smart contracts have been proposed to make contract management easier. Progress in construction payments is crucial for efficient project execution and financial welfare of its partners according to Navon [1996]. The building industry is still using traditional contracts and time-consuming payment systems that rely on manual processes. The project stakeholders suffer as a consequence of late or non-payments Peters et al. [2019], which makes construction industry heavily credit and poor cash sector. For payment stakeholders such as suppliers and sub- contractors, payment issues are much more serious, Kaka [2001].

Smart contracts have been proposed to make contract management easier. Smart contracts are made by analyzing the relationships between contract participants and contractual data, then modeling traditional textual contracts in XML format, Cardoso and Oliveira [2008]. However, Smart contracts are currently used mostly in the IT industry, where the complexity of relationships between parties, obligations, and activities is lower than in building contracts. Smart contract can support automated construction payments, which may benefit Stakeholders in the building supply chain comprising creditors, owners, construction companies, subcontractors, vendors of materials, and rental equipment companies. Savings on costs, times and untimely payment can be achieved by automation, which is an important construction risk factor, with catastrophic impacts, including cost overrun, delays and the re-engaged trust, Durdyev and Hosseini [2019]. This resultant lack of confidence further reduces payment processing and creates an unsafe loop claimed by Manu et al. [2015]. Increased availability of as-built digital information, backed by automation, machine learning and Building Information Modeling (BIM), has been added to today's trend toward automation, Santos et al. [2017], Bilal et al. [2016]. Payment automation, however, is distant from reality because existing payment programs do not provide or allow automation of progress data. In order to address the inefficiency of payment systems, smart block-chain-supported contracts seems promising, Penzes et al. [2018]; however, what is different between the existing means of automation payment opting for current payment solutions, such as computerization is not clear at all. The authors feel that studying why block-chain is critical for its successful use with relation to advancement in building and how. Ad- dressing this basic subject is an important element towards a clearer objective use, driven by industry challenges and not a transitional concern for new technology, of smart contracts.

In the bid to attain this objective, this research first explores the underlying impediments to automation. The study subsequently proposes that, because of its decentralized and assured functionality of contracts, smart block-chain based contracts can tackle the barriers outlined and provide reliability into payment governance. In order to achieve this goal, this study first explores the underlying impediments to automation. The study subsequently proposes that, due to their decentralized and guaranteed execution of contracts, smart block-chain based contracts can tackle the barriers outlined and provide reliability into payment governance.

## 1.1 Research questions:

1. Is it feasible to implement digital payment automation in construction project governance?
2. Can construction contracts be digitized by block-chain applications?

In order to find answers to above stated research question, this study is aimed at finding out the ground facts for the smart contract based construction payment automation implication with the help of block-chain technology, to better understand the application benefits and clear the way for further development in the sector for practical implementation.

# 2 Construction contracts:

Construction contracts offer specific rights in respect of payments to the project participants and 'since the social structure is a right, if the community or group does not acknowledge the right owner's authority, then that right will not continue', Talbot-Jones and Bennett [2019]. Social structures are hence contracts used for compensation of performance. The conception, implementation and enforcement depend on the conventions of individuals and on the collective intentionality of the project members.

In order to define social buildings in view of their underlying legislation, the work under discussion uses the evolution of social reality, Searle [2010]. Searle's work has been inspired by a preoccupation with people's potential to incorporate objectivity into merely social systems. Progress is based on data describing the progress on the worksite, project information, and monetary and operator rights swapping. The study distinguishes between these workflows for the data and information utilized in both categories. This difference relies on the connection of both work and objective observer. Certain facts are based on the human and human consent and are true irrespective of their observers. How do project stakeholders turn basic physical observations into institutional facts? The work performed by subcontractor, for example, must be translated to the fact that the subcontractor has the right to reimbursement. In order to describe the links between facts, Searle created regulatory and constitutive norms. The driving regulations are regulatory because they are not subject to the driving act.

Figure 1(b) illustrates the significance of two supply chain channel of construction contracts that critically affect the applications for payment: product (progress in activity) and cash (payment) payments for progress in the development namely, the social reality and reality correspondingly.

Automation of progress compensation must entail the use of components without human involvement; such computerized contract codes and self-employed workers utilized for the monitoring of work progress (for example, land based monitoring vehicles, aerial drones, or some other remote technologies) without human subjectivity. it's necessary to act as constitutive regulations for the contractual structures employed to bridge the gaps between product flow (physical reality) and cash flow(social reality), figure 1(b).

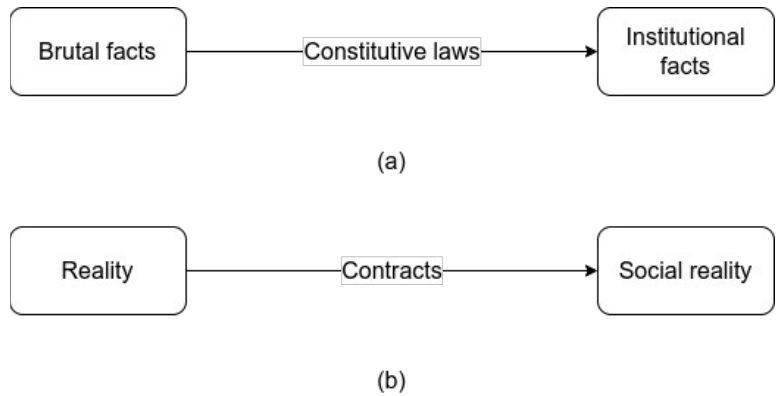

(a)

(b)

Figure 1: Construction contracts: product flow (brute facts) to cash flow (institutional facts)

The authors contend that existing contract documents and payment requests are regulatory merely. They cannot operate as constitutive regulations because of its dependence on centralized system for collecting, modeling and transmission of product and financial rewards and absence of self-realization systems, as a result of centralized mechanisms and guaranteed execution.

## 2.1 Centralized Payment System:

Workflows for the construction interim payments (fig.2) include packed cooperation with competing business objectives between independent players. The workflows must be significantly managed by this un-trustful environment, Bitran et al. [2007]. Payments are prepared, reviewed, approved, implemented and enforced using centrally trusted procedures. In the building supply chain, banks also participated significantly as trusted intermediaries in processing cash flow and transactions in addition to project parties. This high level of centralization has had significant effects which damaged the feasibility of the automation of payments.

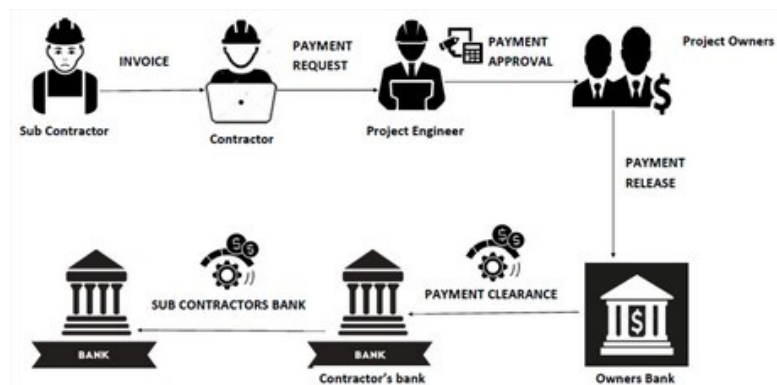

Figure 2: Construction progress payment: traditional payment flow

First of everything, supply chain operations (which is: the product and payment for the product) might be kept in slices among all the stakeholders of the project and financial institutions (fig.2). It prevents integration of the supply chain and further splits product and cash according to Blount [2008]. This center oriented and compartmentalized data collection, and documentation techniques eliminate the ability to produce a trustworthy single source of truth. Consequently, the accuracy of various data fragments must be checked and the transformation from a product flux to a payment realization cannot be automated.

Secondly, in-out payments (cash flow) data are collected, stored and interpreted by external stakeholders (banks), which are different from significant players such as project owners and financial lenders and the contractors. For investment in successful implementation, these external actors have a key position in the supply chain of the building industry, which grants them exclusivity to data that is essential for supply chain, and optimization of process, Silvestro and Lustrato [2014]. This dependence upon the banks affected the supply chain integration, nevertheless a significant step towards filling the gap of the brute facts and institutional reality of a project. It should also be noted that the financial institutions themselves utilize an intermediary network according to Fellenz et al. [2009].

Third, centralization diminishes energy concentration and generates bottlenecks that could stymie payment processing, Manu et al. [2015]. This asymmetry of knowledge is exacerbated by unequal exposure to the information chain, which encourages opportunistic behavior. The resulting lack of confidence prevents payment processes from being automated because parties must constantly verify validity facts, Feldmann and Müller [2003].

## 2.2 Trust Issues in Payment Execution:

It cannot be guaranteed to execute the contractual agreements and condition, as signed by the agreeing parties, causing in delays or non-payments and defaults on the other hand. It is partly because of the form of traditional contractual documents employed, and its dependence on central bodies, for example, the Courts of Justice (CoJ) for contract enforcement, and only after one default. Contractors' unanticipated predictions of their cash flow and working capital remain uncertain as a result of this failure to assure the timing and exact quantity of payments, [PricewaterhouseCoopers]. Over-billing is a result of this fear, as well as contractors' reactions to slow-paying clients who threaten to put them in a negative cash flow situation and lead to insolvency. This uncertainty is particularly troublesome for subcontractors, as payment is frequently contingent not only on the client but also on the general contractor. Despite the fact that all of the project's partners and their stakeholders benefit from the timely reimbursement (Kaka 2001), there is nothing to guarantee payment on the basis of precise contractual conditions. For example, payments to suppliers or subcontractors can also be delayed even after the owner pays to the contractors, because of time-consuming lien waivers or because the contractor uses the fund to tackle its own cash flow concerns.

## 3 Smart Contracts:

Szabo [1994], established the concept for Smart Contracts, describing it as "a system for computer protocols which implements the terms and conditions of an agreement." His definition advises that an automated protocols (1) compliant contractual agreement, (2) malicious and unintentional errors could be minimized and (3) It is possible to eliminate the role of inter-mediators in contract enforcement.

Human interest must be formalized in computerized contracts, which strengthens the security of contractual relations and agreements, he stated in his 1997 book, Szabo [1997a]. Szabo [1997b] suggested that, as compared to paper and traditional contracts, protocols that run on a public network provide a better way to formalize connections.

For smart contracts observability, verification and privacy(protection against any other parties) are the main features as presented by Szabo [1997b]. The observability feature assures that none of the parties is covertly informed prior to the conclusion of a contractual agreement (ex ante) and after (ex post). Parties should be able to decide whether to proceed optimally. The verifiability guarantees that the adjudicator can demonstrate that a party has failed to comply with its terms of contract, based on the formalization of rules and relations via smart contracts. The privacy requirement is that, unless they participate as adjudicators, the smart contract must be contained by third parties.

Although Szabo [1994] discussed some possible applications of digital contracts, such as the intelligent real estate, The idea failed to acquire momentum because of the limited real-world implementation of smart contracts; it was Bitcoin's inventiveness, Nakamoto [2008] and the emergence of a block-chain that led to smart contracts once again in 2010.

# 4 Blockchain:

Block-chain one of the financial building blocks,is the key reason for cryptocurrencies like Bitcoin's recent success, Nakamoto [2008]. It offers a decentralized approach for achieving peer-to-peer network consensus (P2P), Narayanan et al. [2016]. It utilizes encryption and game theory tools to develop and maintain an ingenious and distributed common directory of transactions without Banks, for example, are centralizing authority, Swan [2015]. The block-chain decentralized consensual mechanism does away with the requirement for transaction participants to have faith in each other; many experts consider this to be the most important trait that distinguishes cryptocurrencies from their failed predecessors, claimed by Vigna and Casey [2016].

A blockchain includes a sequence of data blocks, each having a crypto-graphical summary of its preceding block, storing many transactions. As a result, the record chain is immutable and traceable. Introducing the concept of block-chains in decentralized applications the Ethereum protocol took this concept farther, Wood and others [2014]. Ethereum is a cryptocurrency with its own block-chain (Ether). It is considered version 2.0 of the Block-chain and goes far beyond decentralized currency exchange, Diedrich [2016]. On the Ethereum virtual computer (EVM), the protocol and nearly the whole language enable for the program execution, which is a decentralized global computer. This computer network executes the code and publishes for the underlying block-chain; it provides a consensus-based approach similar to Bitcoin. The global computer's decentralized structure eliminates manipulation of the contracts and minimizes delay.

# 5   Smart Contracts for Construction Projects:

For Distributed Ledger Technology (DLT) and block-chain, Construction administration was among the selected research areas, Li et al. [2019a]. To investigate the benefits of smart contracts in the construction industry, focus groups were used to develop sociol-technical frameworks, Li et al. [2019b]. For example, decentralized digital contracts in the building industry are expected to decrease costs by 9%, Dakhli et al. [2019]. One of six possible application areas for intelligent contracts and block-chain has been recognized as Token-based payment systems, Hunhevicz and Hall [2019].

These discoveries have encouraged the attention on the ability of constructing information modeling for a block-chain and smart contract, by boosting finance monitoring and raising the credibility of project data, Block-chain may supplement the current techniques to modeling of information, claimed in a research by Turk and Klinc [2017].

Smart contracts were shown to be an extension of BIM methods in a research, Mason [2019], whereas, in a new study, Gabert and Grönlund [2018], such integration was not regarded an imminent research topic. Others talked about arguments to and against it, Mason [2019]. Smart contracts were also found to be unsuitable for complex construction projects in an industry survey, Gabert and Grönlund [2018]. Other researchers showed possible advantages for post-disaster recovery via BIM and blockchain, Nawari and Ravindran [2019]. Building information content adjustable, immutable and distributed perspective of the Building Information System, Hamledari and Fischer [2021].

In spite of the possible applications revealed in the studies, adoption success necessitates a deeper knowledge of the links between industry challenges and the essential aspects of smart contracts, Hunhevicz and Hall [2020]. In terms of the automation in interim and process of supply chain, although block-chain based smart contracts have the potential to deliver on their expectations. It's unclear what distinguishes this technology from a computerized payment systems or other possible kinds of automated payment, Yang et al. [2020].

In the context of payment mechanization, it is crucial to examine the underlying barriers to automation and their relationship to the distinguishing features of block-chain and programmable contracts, which does not exist in the literature.

# 6   Decentralization of Smart Contracts:

Even if it is automated, contract documents cannot facilitate efficient advance payment automation. This is due to two main limitations: firstly the dependence on centralized execution, control, modeling and communication methods for the flow; and the lack of assured contractual execution. The establishment of components that act as constituent rules is required for effective advance payment automation. Because of its two main properties, smart block-chain-enabled projects can solve this demand. Firstly, smart contracts decentralize payment execution and cause havoc with payment management. Second, it ensures the contract provisions are assured to be executed and the unforeseen results for the parties are eliminated.

These characteristics set smart contracts apart from alternative technologies that could be utilized to allow self-payment. To emphasize this notion, Fig.-3 compares the system architecture of two of the automated payment applications that operate without and with smart contracts. Contract documents are digitized and converted to executable in both situations, which settles payment amounts based on input progress data. These two automation modes are contrasted in the next sections.

## 6.1  Centralized control automation:

One Project Member is required to implement the coding for the contract and distribute calculation outcomes to remaining P2P Network of project fig.-3(a) for a system functioning without smart contract. Subsequently, banks and financial institutions need to process the monetary transactions further and use the results of the information of accounts which are to be paid or received. This results in client- server architecture with a single player centralizing code executions, authentication, and permissions (e.g. the owner). The present contract documents are governed by the same limitations: which are the lack of its independence and centralized controlling. These servers has the able in refusing services to other stakeholders, delay some processes (e.g. delay transactions which a supplier considers), and amend the contract code underlying the project and rewrite historical data including built-in BIM and acquired progressive data. This centralized implementation of the code causes failures, elevates downtime and raises the asymmetry of knowledge. Such automatic payment is not trustworthy.
"Those who control the past control the future," the central control of the server over the execution of current transactions and the option of re-writing prior projects skew the opportunistic conduct and power centralization. This architecture builds significantly on the confidence of stakeholders in a trustless environment where business targets compete. This is not an automated payment option.

## 6.2  Decentralized automation of smart contracts:

On the other hand, the digitized contract will be implemented and executed in digitized contract-based settings Fig.-3(b) on an open block-chain rather than in the P2P project network. For example the Ethereum Virtual Machine (EVM) fig.-3(b) behaves as a fault-tolerant global workstation for the Ethereum block-chain, in which each node implements the smart contract and publishes the simulation outputs to the ledgers' bottom line (i.e., Ethereum block-chain). The project stakeholders in have no control over contract implementation once they have been implemented, nor can they change or end the administration of payments. This is a decentralized control mechanism. The computer-based contract, the smart public block-chain con- tract, is guaranteed to be executed. This disintermediation benefits from a stronger integration of the product flow and cash flow on the same platform (in ledgers), enhanced supply chain integration, increased visibility of the supply chain and a reduction in the commission charges applied by external institutions for transactions and minimized processing time. However, this middle ground technique introduces a new difficulty for the P2P network, namely the reality because money is a notion of society, and depends upon financial institutions to operate; it now has an inter- country role and is not engaged immediately with the flux of cash between parts of the supply chain. The block-chain based system has been resurrected because it allows for a periodic interim transfer of payments through cryptocurrencies and provides protocols for unique crypto-asset design. The importance of decentralization and assured execution is demonstrated by the contrast among both the smart architecture and customer server, the essential features that elaborate and make them constitutive by defining the role of intelligent agreements in automated payment systems. Its reliability comes at the expense of inefficiency, which block-chain offers. Only the server is responsible for the calculations in the decentralized architecture, whereas other EVM nodes execute the same computerized contract and keep a complete copy of the registry and the associated transactions. It provides a secure way for executing electronic contracts without depending on other parties.

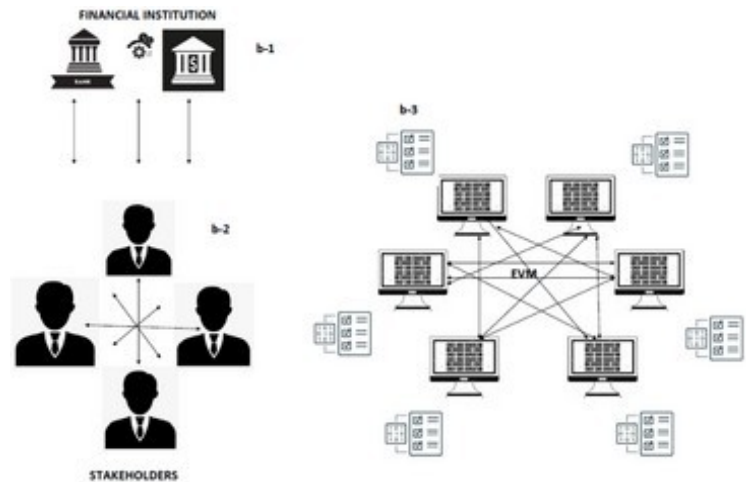

Figure 3: smart contract based Payment system a) without blockchain b) with blockchain, Hamledari and Fischer [2021]

## 7 Implications:

Smart contracts and their underlying block-chain technology have broader implications for computer science, resolution of conflicts, and contract conduct in constructions regulations, in addition to its critical role in automating advances. The primary consequences are that stakeholder cooperation will have more autonomy and will require less trust and an application-independent, life-cyclical and shared perspective on the information about the project - a single source of truth.
The rules of smart contracts are hard-coded into the system and assure stakeholder compliance through their own enforceability characteristics. This decreases the de- pendency on trust and promotes the design of technologies to encourage cooperation between project stakeholders concluding a contractual agreement. The ensuing in- dependence, enabled by secure protocols, minimize the need to manually verify data accuracy, which shows a reduction in trust necessity, claimed by Gad et al. [2016]. Autonomous contract management and Decentralization is more than simply corporate regulator, it is actually a factor in decreasing friction drastically and making conductivity more efficient, as this decreases unforeseen output, De Filippi [2018]. Each project participant inevitably agrees to the contractual conditions as a souvenir in digital agreement that controls project delivery, Abdul-Malak and Hamie [2019]. Contract management autonomy, along with smart contract legacy, provides an effective enforcement mechanism for ensuring compliance with standards and best practices.

Without relying on intermediaries, block-chain and cryptographic protocols are able to obtain agreement without relying on intermediaries on the actual status of a shared reality/content curated collaboratively. Automating progress payments necessitates a shared understanding between the parties engaged on brute or institutional events such as construction progress, as-built payment status, accounts payable and receivable, contract conditions, contractual agreement compliance, and financial flow between the parties. ADR (Alternative dispute resolution) proceedings depends on information currently documented in several versions and in an independent stakeholder managed case law. This inadmissibility of certain evidence none of the court of law creates problems concerning the authenticity of proof. By introducing a new data communication and access mechanism, Block-chain promotes ADR, in which Information is securely saved and written on a common data layer that may be accessed via separate applications.

# 8 Limitations:

- Once a project has been implemented, the decentralized system not only affects the workflows for payments but also the role of stakeholders. A failure to recognize this evolving dynamic presents a significant risk to the adoption of smart contracts, Piao [2020].

- Computational outcomes are not reversible, this implies that bad coding can jeopardize the integrity of project information or result in huge financial losses. The stakeholders must ensure that the computerized code is sound before deploying it. To ensure the operation of an intelligent contract, the academic community has focused on verification and validation approaches to enhance the safety of smart contracts, Bhargavan et al. [2016].

- Regarding security of contracts, other hazards emerge because block-chains are susceptible to P2P assaults. In order to disrupt the consensus of block-chain, a 51% attack on the network is necessary. This involves controlling over than 50% of the hash rate and computational capacity in block-chain utilizing the proof of work protocol (PoW), either by acquisition of other mining pools or through gathering computer power equivalent to the real hash rates and contributing as a miner, Gervais et al. [2016]. Although it cannot guarantee an effective strike, it does increase the chances of causing damage to the consensus process, such as allowing defective data being sent to the shared ledger or ignoring some transactions.

# 9 Conclusion:

The construction sector as of now relies on slow and inefficient traditional payment applications. Projects and parties involved are exposed in this way towards multitude hazards, as well delayed or non-paid transactions, possibility of lien, diminished confidence and overrunning costs/program. The payment automation is still not a reality despite promised advantages and the abundance of digital progress data available now.

In the early stages of research and development, smart contract-based payment systems still exist. Successful adoption of the contract requires: the significance of intelligent deal security and verification, the assessment of the effect on incentive frameworks, and the characteristics of decentralized contract administration project fulfillment methods.

In order to identify the constraints that inhibit automation. The argument was made that existing applications for payment operations and associated documents for contract could not support progress payments automatically, even if they are computerized, due to two important restrictions: The reliance on central methods of control and implementation, as well as the absence of an assured operation, are both issues. The paper then went on to illustrate why, despite the difficulties, block-chain and intelligent contracts offer a lot of promise. Payment automation has been investigated in both block-chain-based and non-block-chain-based systems. The introduction of zero-trust computing owing to the usage of autonomous and automated code is one of the overarching consequences of smart contracts and their underpinning block-chain technology in settling disputes and contractual compliance; and the establishment of life-cyclical, application independent, immutable and fully audited project information. First, by implementing mechanisms that nurture confidence and decrease the potential for opportunistic behavior, and, second, by establishing a single source of reality that can be accessed by all stakeholders, it promotes dispute settlement.